\documentclass{article}

\usepackage{graphicx}
\usepackage{color}
\usepackage{amsmath}
\usepackage{amssymb}

\begin{document}

 \large

\newcommand{\al}{\mbox{$\alpha$}}
\newcommand{\be}{\mbox{$\beta$}}
\newcommand{\ep}{\mbox{$\epsilon$}}
\newcommand{\gam}{\mbox{$\gamma$}}
\newcommand{\sig}{\mbox{$\sigma$}}

\DeclareRobustCommand{\FIN}{%
  \ifmmode 
  \else \leavevmode\unskip\penalty9999 \hbox{}\nobreak\hfill
  \fi
  $\bullet$ \vspace{5mm}}

\newcommand{\calA}{\mbox{${\cal A}$}}
\newcommand{\calB}{\mbox{${\cal B}$}}
\newcommand{\calC}{\mbox{${\cal C}$}}

\newcommand{\muas}{\mbox{$\mu$-a.s.}}
\newcommand{\Nat}{\mbox{${I\!\!N}$}}
\newcommand{\Rea}{\mbox{${\mathbb R}$}}
\newcommand{\Hil}{\mbox{${I\!\!H}$}}
\newcommand{\Prob}{\mbox{${ P}$}}

\newcommand{\nin}{\mbox{$n \in {I\!\!N}$}}
\newcommand{\suc}{\mbox{$\{X_{n}\}$}}
\newcommand{\sucP}{\mbox{$\{P_{n}\}$}}

\newcommand{\conv}{\rightarrow}
\newcommand{\convn}{\rightarrow_{n\rightarrow \infty}}
\newcommand{\convp}{\rightarrow_{\mbox{c.p.}}}
\newcommand{\convs}{\rightarrow_{\mbox{a.s.}}}
\newcommand{\convw}{\rightarrow_w}
\newcommand{\convd}{\stackrel{\cal D}{\rightarrow}}

\newtheorem {Prop}{Proposition} [section]
 \newtheorem {Lemm}[Prop] {Lemma}
 \newtheorem {Theo}[Prop]{Theorem}
 \newtheorem {Coro}[Prop] {Corollary}
 \newtheorem {Nota}{Remark}[Prop]
 \newtheorem {Ejem}[Prop] {Example}
 \newtheorem {Defi}[Prop]{Definition}
 \newtheorem {Figu}[Prop]{Figure}
 \newtheorem {Tabla}[Prop]{Table}

\title{\sc The Random Tukey Depth\footnote{Research partially supported by the
Spanish Ministerio de Ciencia y Tecnolog\'{\i}a, grant
MTM2005-08519-C02-02 and the Consejer\'{\i}a de Educaci\'on y Cultura de la Junta de Castilla y Le\'on, grant PAPIJCL VA102/06.}}

\author{J.A. Cuesta-Albertos and A. Nieto-Reyes\\Departamento de
Matem\'{a}ticas, Estad\'{\i}stica y Computaci\'{o}n,\\
Universidad de Cantabria, Spain}
\maketitle

 \begin{abstract}
The computation of the  Tukey depth, also called halfspace depth, is
very demanding, even in low dimensional spaces, because it requires
the consideration of all possible one-dimensional projections. In
this paper we propose a random depth which approximates the Tukey
depth. It only takes into account a finite number of one-dimensional
projections which are chosen at random. Thus, this random depth
requires a very small computation time even in high dimensional
spaces. Moreover, it is easily extended to cover the functional
framework.

We present some simulations indicating how many projections should
be considered depending on the sample size and on the dimension of
the sample space. We also compare this depth with some others
proposed in the literature. It is noteworthy that the random depth,
based on a very low number of projections,  obtains results very
similar to those obtained with other depths.

 \end{abstract}

\vspace{3mm}


\noindent{\em A.M.S. 1980 subject classification:} Primary 62H05; Secondary: 62G07, 62G35.

\vspace{1.5cm}

\section{Introduction}
This paper is written in the same spirit as \cite{Hand}. In the
abstract of this paper, D.J. Hand states that {\it ``...simple
methods typically yield performance almost as good as more
sophisticated methods to the extent that the difference in
performance may be swamped by other sources of uncertainty..."}.
Hand's work is related to classification techniques. Here we analyze
a conceptually simple and easy to compute multidimensional depth
that can be applied to functional problems and that provides results
comparable to those obtained with more involved depths.

Depths are intended to order a given set in the sense that if a
datum is moved toward the center of the data cloud, then its depth
increases and if the datum is moved toward the outside, then its
depth decreases.

More generally, given a probability distribution \Prob \ defined in
a multidimensional (or even infinite-dimensional) space $\cal X$, a
depth tries to order the points in $\cal X$  from the ``center (of
\Prob)" to the ``outward (of \Prob)". Obviously, this problem
includes data sets if we consider \Prob \ as the empirical
distribution associated to the data set at hand. Thus, in what
follows, we will always refer to the depth associated to a
probability distribution \Prob .

In the one-dimensional case, it is reasonable to  order the points
using the order induced by the function
\begin{equation}Ê\label{Depth1}
x  \rightarrow D_1(x,\Prob ) := \min \{ \Prob (-\infty,x],\Prob[x,\infty)\}.
\end{equation}
Thus, the points are ordered following the decreasing order of the
absolute values of the differences between their percentiles and 50,
and the deepest points are the medians of $\Prob$.

Several multidimensional depths have been proposed (see, for
instance, the recent book \cite{Dimacs}) but here we are mainly
interested in the {\it Tukey  (or halfspace) depth} (see
\cite{Tukey}).
If $x \in \Rea^p$, then, the Tukey depth of $x$ with respect to $\Prob, \
D_T(x,\Prob)$, is the minimal probability which can be attained in
the closed halfspaces containing $x$. According to \cite{Zuo00},  this  depth  behaves very well in comparison with various competitors.

An equivalent definition of $D_T(x,\Prob)$ is the following. Given $v \in \Rea^p$, let  $\Pi_v$ be  the projection of $\Rea^p$ on the one dimensional subspace generated by $v$. Thus,  $\Prob  \circ \Pi_v^{-1}$ is the marginal of \Prob \ on this subspace, and it is obvious that
\begin{equation} \label{TukeyDepth}
D_T(x,\Prob)=\inf \{ D_1(\Pi_v(x),\Prob  \circ \Pi_v^{-1} ) : v \in \Rea^p \}, \ x \in \Rea^p.
\end{equation}
I.e., $D_T(x,\Prob)$ is the infimum of all possible one-dimensional
depths of the one-dimensional projections of $x$, where those depths are computed
with respect to the corresponding (one-dimensional) marginals of
\Prob .  Some other depths based on the consideration of all
possible one-dimensional projections, but replacing $D_1(x,\Prob )$
by some other function, have been proposed (see, for instance, \cite
{Zuo03}). We consider that what follows could be applied to all of
them, but, we have chosen the Tukey depth to test it concretely.

Perhaps the most important drawback of the Tukey depth is the
required computational time. This time  is more or less reasonable
if $p= 2$, but it becomes prohibitive even for $p= 8$ \cite[pag.
54]{Mosler}. To reduce the time, in \cite{Zuo06} (page 2234) it is
proposed to approximate their values using randomly selected
projections.

On the other hand, in \cite{Cuevas07}, a random depth is defined. In
this paper, given a point $x$, the authors propose to choose at
random a finite number of vectors $v_1,...,v_k$, and then, take as
depth of $x$ the mean of the values $ D_1(\Pi_{v_i}(x),\Prob \circ
\Pi_{v_i}^{-1} ), i=1,...,k$.

Our approach follows more closely the suggestion in   \cite{Zuo06}:
We simply replace the infimum in (\ref{TukeyDepth}) by a minimum
over a finite number of randomly chosen projections.

\begin{Defi} \label{DefiRT}
Let \Prob \ be a probability distribution on $\Rea^p$. Let $x\in
\Rea^p$, $k\in \Nat$ and let $\nu$ be an absolutely continuous
distribution on $\Rea^p$. The random Tukey depth of $x$ with respect
to \Prob \ based on $k$ random vectors chosen with $\nu$ is
\[
D_{T,k,\nu}(x,\Prob) = \min \{ D_1( \Pi_{v_i}(x), \Prob  \circ \Pi_{v_i}^{-1}) : i=1,...,k \}, \ x \in \Rea^p,
\]
where $v_1,...,v_k$ are independent and identically distributed
random vectors with distribution $\nu$.
\end{Defi}

Obviously, $D_{T,k,\nu}(x,\Prob) $ is a random variable. It may seem
a bit strange to take a random quantity to measure the depth of a
point, which is inherently not-random. We have two reasons to take
this point of view.

Firstly, Theorem 4.1 in \cite{Cues07} shows  that if $P$ and $Q$ are probability distributions on
$\Rea^p$, $\nu$ is an absolutely continuous distribution on $\Rea^p$
and
\[
\nu \{v\in \Rea^p: P\circ \Pi_v^{-1} = Q\circ \Pi_v^{-1} \} >0,
\]
then $P=Q$. In other words, if we have two different distributions,
and we  randomly choose a marginal of them, those marginals are
almost surely different. In fact, it is also  required that at least
one of the distributions is determined by their moments, but this is
not too important for the time being. According to this result, one
randomly chosen projection is enough to distinguish between two
$p$-dimensional distributions. Since the depths determine
one-dimensional distributions, a depth computed on just one random
projection allows to distinguish between two distributions.

Secondly, if  the support of $\nu$ is $\Rea^p$, and, for every $k$, $\{v_1,...,v_k\}Ê\subset \{v_1,...,v_{k+1}\}$, then
\begin{equation} \label{Eqas}
D_{T,k,\nu}(x,\Prob) \geq D_{T,k+1,\nu}(x,\Prob)
\rightarrow D_{T}(x,\Prob), \ \mbox{ a.s.}
\end{equation}
Therefore, if we choose a large enough $k,$ the effect of the
randomness in $D_{T,k,\nu}$ will be negligible. Of course, the
question of interest here is to learn how large $k$ must be, because
values of $k$ that are too large would make this definition useless.

One way to select $k$ is to compare $D_T$ and $D_{T,k,\nu}$, but the
long computation times required to obtain $D_T$ make those
comparisons unpractical. Instead of this, we have decided to choose
a situation in which the deepness of the points are clearly defined
and can easily be computed with a different depth.

If  \Prob \ is  an elliptical distribution
with centralization parameter $\mu$ and dispersion matrix $\Sigma$, then,
it seems that every reasonable depth should consider $\mu$ as the
deepest point, that points at the same Mahalanobis distance of $\mu$  should have the
same depth, and that differences in depth should correspond with differences in Mahalanobis distance of $\mu$. Then, in this
situation, every depth should be a monotone function of the
Mahalanobis depth \cite{Maha36}, where, given $x\in \Rea^p,$ this
depth is
\begin{equation} \label{EdDM}
D_M(x,\Prob) :=\frac 1{1+(x-\mu)^t\Sigma^{-1}(x-\mu)}.
\end{equation}

Therefore, we can choose the right $k$ in $D_{T,k,\nu}$ as follows:
If \Prob \ is elliptical, $D_T(\cdot ,\Prob)$, is a monotone function of
$D_M(\cdot ,\Prob)$. Thus, from (\ref{Eqas}), the larger the $k$, the
larger the resemblance between $D_{T,k,\nu}(\cdot,\Prob ) $ and a
monotone function of $D_{M}(\cdot,\Prob ) $. However, there should
exist a value $k_0$ from which this resemblance starts to stabilize.
This is the value for $k$ we are looking for.

However, in practice, we do not know $P,$ and we only have a random
sample. It seems that  the selection of $k_0$ should take this fact
into consideration. In Section \ref{SecElegirk} we present a
procedure to do this.

The results of the comparison of $D_{M}(\cdot,\Prob)$ and
$D_{T,k,\nu}(\cdot,\Prob)$ for several sample sizes, dimensions and
elliptical distributions are shown in Table \ref{TablaValoresk2}.
According to this table $k=36$ is the maximum number of directions
required if the sample size is below 1,000.

Once the right values of $k$ have been fixed, we carry out, also in Section \ref{SecElegirk}, a study to compare $D_{T,k,\nu}$ with $D_{T}$ from the applications point of view. The results are quite encouraging.

Section \ref{SecElegirk}  ends with a comparison of the time
required to compute $D_{T,k,\nu}$ and the time to compute $D_{M}$.
This comparison  turns out to be favorable for $D_{T,k,\nu}$.

An important advantage of Definition \ref{DefiRT} is that it can be applied in every space in which projections can be computed. Since this is an easy task in Hilbert spaces and Theorem 4.1 in  \cite{Cues07} holds in separable Hilbert spaces, we  propose to employ Definition \ref{DefiRT} to compute depths of points in those spaces.

A difference with the $p$-dimensional case is that here we are not
aware of any situation  in which a gold standard to compare depths
does exist. However, in \cite{Romo06}, authors employ functional
depths in a classification problem. In Section \ref{SecFunctional},
we compare the results obtained with the random depth with those
obtained in \cite{Romo06} in the same problem.

This study reinforces the feeling that the values obtained in Table
\ref{TablaValoresk2} are accurate. Following this table, in Section
\ref{SecFunctional},  we
have taken $k=10$ because the sample sizes are around 50. The results have been
satisfactory even if there is no reason to assume any particular
model on the distribution generating the samples.

Some other functional depths (not considered here) have been
proposed in the literature. We are aware of the Fraiman-Muniz depth
(introduced  in \cite{Frai}), the $h$-mode depth (proposed in
\cite{Cuevas06}) and the above mentioned random depth and a double
random depth (RPD) which appear in \cite{Cuevas07}. An interesting
application of those depths to outlier detection is made in
\cite{Febre07}.

In \cite{Cuevas07}, the authors apply the depths that they analyze
to the same classification  problem that we study here. The
proportions of right classifications that they obtain with depths
are similar to those reported here except for the RPD depth. This is
a random depth which takes into account not only the curves but also
their derivatives. Thus, it handles more information than we employ
here, and the results are not comparable.

We want to mention that Theorem 4.1 in \cite{Cues07}  provides the
theoretical background for the random depth proposed in
\cite{Cuevas07} whose definition, in fact, only considers a vector.
The  only reason the authors give for handling $k$ ($ > 1$) randomly
chosen vectors is to provide more stability to the definition.
Moreover, Theorem 4.1 in \cite{Cues07}  has also been applied to
construct goodness of fit test, for instance, in \cite{Cues07b},
\cite{Cues07c} and \cite{Cues06}.  In those papers, the authors also
handle more than one projection. They take $k$ ranging from 1 to 25
in  \cite{Cues07b}, ranging from 2 to 40 in   \cite{Cues07c}, and
$k=100$ in \cite{Cues06} with the same objective as in
\cite{Cuevas07} and also with no specific reason to make those
selections.

We consider that the results provided in this paper could help to settle the way in which the number of random projections should be chosen.

Computations have been carried out with MatLab. Computational codes are available from the authors upon request.

\section{How many random projections? Testing homogeneity} \label{SecElegirk}
In this section we analyze the $p$-dimensional case. Obviously,
Theorem 4.1 in \cite{Cues07} also holds if we take $\nu$ a
probability distribution absolutely continuous with respect to the
surface measure on the unit sphere in $\Rea^p$. We are also
interested in what (\ref{Eqas}) holds. Then, in this section,  we
fix $\nu$ to be the uniform distribution on the unit sphere, and we
will suppress the subindex $\nu$ in the notation $D_{T,k}$.

As stated in the introduction,  to decide how to choose $k$, we will analyze the case in which \Prob \ is an elliptical distribution by comparing the functions $D_M(\cdot,\Prob )$ and $D_{T,k}(\cdot,\Prob )$ for several values of $k$.

Taking into account that depths only try to rank points according to
their closeness to the center of \Prob , it is reasonable to measure
the resemblance between  $D_{T,k}(\cdot,\Prob ) $ and
$D_{M}(\cdot,\Prob ) $ looking only at the ranks of the points. This
is equivalent to employing the Spearman correlation coefficient,
$\rho$. Thus, the resemblance that we handle here is
  \begin{equation}\label{EqSpearman}
r_{k,P }: =\rho \left( D_{T,k}(X,\Prob ),D_{M}(X,\Prob ) \right) ,
\end{equation}
 where $X$ is a random variable with distribution \Prob .

If \Prob \ is an elliptical distribution, then the function $k \conv
r_{k,P }$ is strictly increasing. We try to identify $k$ with the
point $k_0$ from which the increments become negligible.

Moreover, in practice, we will not have a distribution \Prob , but a
random sample $x_1,...,x_n$ taken from \Prob . This leads us to
replace \Prob \ in (\ref{EqSpearman}) by the empirical distribution
$\Prob_n$ ($\Prob_n [A]Ê= \#(A \cap \{x_1,...,x_n\})/n$) which does
not follow exactly the model and, consequently,  the function
$r_{k,P _n}$ is not necessarily increasing. We propose is to
identify $k_0$ with the point in which $r_{k,P _n}$ starts to
oscillate or, more precisely, estimate $k_0$ by
\[
\hat k_0 = \infÊ\{ k \ge 1 :  r_{k,P _n} > r_{k+1,P _n} \}.
\]

To check the dependence between $\hat k_0$ and the underlying
distribution, we employ samples taken from multidimensional standard
Gaussian distributions, from distributions  with independent double
exponential marginals and with independent  Cauchy marginals. We are
also interested in looking at the dependence between $\hat k_0$ and
the dimension of the space and the sample size. To do this, we have
selected five dimensions ($p=2, 4,8,25, 50$),  and six sample sizes
($n=25 , 50 , 100 , 250 , 500, 1,000$).

We need to compute the location center and the dispersion matrix of
$\Prob_n$ to be employed in $D_M$. Those parameters should depend on
the distribution which have generated the sample. We mean the
following: the covariance matrix is an appropriate parameter in the
Gaussian and exponential case. But it is not adequate for the Cauchy
distribution, where, we have identified $\Sigma$ with the robust
covariance matrix proposed in \cite{Maronna}, page 206. On the other
hand, we have replaced $\mu$ by the sample mean in the Gaussian case
and by the coordinate-wise median in the exponential and Cauchy
settings.

We have done 10,000 simulations under each set of conditions. In
Table \ref{TablaValoresk2}  we show the mean and the 95\% percentile
of the values obtained for $\hat k_0$.

\begin{Tabla} \label{TablaValoresk2}
Mean and 95\% percentile of the optimum values for the number of
required random projections $k$ for the sample sizes, dimensions and
distributions shown.

Symbol * means that Mahalanobis depth is not defined in those cases
because the dispersion matrix is degenerated.

\end{Tabla}

\begin{tabular}{ccr|cccccc}
&&& \multicolumn{6}{c}{Sample sizes}
\\
[-2mm]
Dimension & Distribution &&&&& &
\\
[-3mm]
 & &   & 25 & 50 & 100 & 250 & 500 &1,000
\\
\hline
\hline
$p=2$ & Gaussian & mean &
 3.61    &    3.90    &    4.17    &    4.19    &    4.23    &    4.20
\\
& &95\% percentile &  6    &     7    &     8    &     8    &     8    &     8
\\
\cline{2-9}
& D. Expone. & mean &
 3.43ÊÊÊ & ÊÊÊ 3.77ÊÊÊ & ÊÊÊ
3.97ÊÊÊ & ÊÊÊ 4.18ÊÊÊ & ÊÊÊ 4.33ÊÊÊÊ & ÊÊÊ 4.44
\\
& &95\% percentile & 6ÊÊÊÊ & ÊÊÊ 7ÊÊÊÊ & ÊÊÊ 7ÊÊÊÊ & ÊÊÊ 8ÊÊÊÊ & ÊÊÊ
9ÊÊ    &   9
\\
\cline{2-9}
& Cauchy& mean & 3.25    &   3.56    &   3.81    & 4.06
&   4.36    &   4.82
\\
& &95\% percentile &  6    &    6    &    7    &    8    &    9 & 11
\\
\hline
\hline
$p=4$ & Gaussian & mean &
4.06    &    5.53    &    7.19    &    9.12    &    9.99    &   10.61
\\
& &95\% percentile &          7    &    10    &    12    &    16    &    18    &    20
\\
\cline{2-9}
& D. Expone. & mean &
 3.95ÊÊÊ & ÊÊÊ 5.25ÊÊÊ & ÊÊÊ 6.69ÊÊÊ & ÊÊÊ 8.49ÊÊ & ÊÊÊ 9.30 Ê&Ê 9.93
\\
& &95\% percentile & 7ÊÊÊÊ & ÊÊÊ 9ÊÊÊ & ÊÊÊ 12ÊÊÊ & ÊÊÊ 15ÊÊÊ & ÊÊÊ
17Ê&  19
\\
\cline{2-9}
& Cauchy& mean &
   3.55    &   4.44    &   5.48    &   6.66    &   7.43    &   8.00
\\
& &95\% percentile & 6    &    8    &   10    &   12    &   14    &
16
\\
\hline
\hline
$p=8$ & Gaussian & mean    &   3.45    &    4.80    &    6.91    &   11.56    &   15.80    &   20.18
\\
& &95\% percentile & 6    &     9    &    13    &    20    &    27    &    35
\\
\cline{2-9}
& D. Expone. & mean &
 Ê 3.76ÊÊÊ & ÊÊÊ 5.11ÊÊÊ & ÊÊÊ
7.53 Ê&Ê 11.98Ê&Ê 15.80ÊÊÊ & ÊÊÊ 19.41
\\
& &95\% percentile & 7ÊÊÊÊ & ÊÊÊ 9ÊÊÊ & ÊÊÊ 13ÊÊÊ & ÊÊÊ 20ÊÊÊ & ÊÊÊ
27Ê&  34
\\
\cline{2-9}
& Cauchy& mean & 3.58   &    4.64   &    6.23   & 8.53
& 10.06   &   11.43
\\
& &95\% percentile & 6   &     9   &    11   &    16   &    19   &    23
\\
\hline
\hline
$p=25$ & Gaussian & mean & *&  3.05    &   4.25    &   6.53      &    10.07      &    16.04
\\
& &95\% percentile&   * & 6    &    9    &   14    &   20    &   29
\\
\cline{2-9}
& D. Expone. & mean &
 *  &  3.84ÊÊÊ & ÊÊÊ 5.27ÊÊÊ & ÊÊÊ 8.93ÊÊ& 14.08ÊÊÊÊ & ÊÊÊ 21.64
\\
& &95\% percentile & *  & 7ÊÊÊ & ÊÊÊ 10ÊÊÊ & ÊÊÊ 17ÊÊÊ & ÊÊÊ 25Ê& Ê
36
\\
\cline{2-9}
& Cauchy& mean &
 * & 4.74   &    6.67   &   10.22   &   13.54   &   16.50
\\
& &95\% percentile &           * &  9   &    12   &    18   &    24
& 30
\\
\hline
\hline
$p=50$ & Gaussian & mean  &    * &   * & 3.16    &   4.71    &   6.55    &   9.96
\\
& &95\% percentile&    * &   * &    7    &   10    &   14    &   20
\\
\cline{2-9}
& D. Expone. & mean &
     * &     * &  4.17ÊÊÊ & ÊÊÊ 6.48ÊÊÊ & ÊÊÊ 9.88ÊÊÊ & ÊÊÊ 15.64
\\
& &95\% percentile &                * &     * & 8ÊÊÊ & ÊÊÊ 13ÊÊÊ &
ÊÊÊ 19Ê& Ê 28
\\
\cline{2-9}
& Cauchy& mean &
     * &     * & 6.75   & 10.94 & 15.16 & 19.61
\\
& &95\% percentile &                * &     * & 12   & 19  & 26  &
34
\\
\hline
\hline
\end{tabular}

\vspace{5mm}

Since, in each simulation, the obtained value of $\hat k_0$ is
bounded from below by 1 and it can take arbitrary large values, the
distribution of $\hat k_0$  is right skew. Thus, the mean produces
larger values than the median giving some guaranty against the
possibility to selecting  values that are too low. Moreover, even if
the mean could be a reasonable selection, we chose the 95\%
percentile for additional safety.

It is apparent from Table \ref{TablaValoresk2} that, in every
dimension, the optimum value for $k$ increases with the sample size.
This increment  is due to the fact that when $n$ increases, the
function $r_{k,P_n }$ approaches $r_{k,P }$, which is strictly
increasing. In other words, when we take a random sample, the
randomness introduces some noise in the model which makes taking
high values of $k$ useless.  However, when $n$ increases, this
randomness is lower,  and, then, it is worth it to increase $k$.

The variation of the optimum $k$ with the dimension is more
striking. If we fix a sample size, in the Gaussian and exponential
case, it happens that the optimum value, first increases with $p$
but, after a change point, it decreases. However, the change point
increases with the sample size.

It seems obvious that the number of projections required to
 accurately represent a cloud of points should increase  with $p$.
But, why does this number decrease after the change point? The answer
lies in the noise introduced by the randomness in taking the sample.
The problem is that we are comparing the random Tukey depth with the
Mahalanobis one. But, in order to compute $D_M$ we need to estimate
the dispersion matrix and, if we keep the sample size fixed, this
estimation worsens  when the dimension increases. The noise
introduced by this fact makes considering high values for $k$
useless. We will briefly analyze this point.

Let us consider the Gaussian case. In order to figure out  how good
the estimator sample covariance is depending on the sample size and
the dimension of the space, we show in Table \ref{TableCovar} the
mean of the determinants of the sample covariance matrices for the
same sample sizes and dimensions as in Table \ref{TablaValoresk2}
obtained along 1,000 simulations taken from a standard Gaussian
distribution.

\begin{Tabla} \label{TableCovar}
Mean of the determinants of the sample covariance matrices computed
in 1,000 random samples taken from a standard Gaussian
distribution.\end{Tabla}

\begin{center}
\begin{tabular}{c|ccccccc}
& \multicolumn{7}{c}{Sample sizes}
\\
[-2mm]
Dimension &  &&&&&&
\\
[-3mm]
 &      25 &  50 & 100 & 250 & 500 &1,000 
\\
\hline
$p=2$
&.953    ÊÊÊ & ÊÊÊ .982    ÊÊÊ & ÊÊÊ .990    ÊÊÊ & ÊÊÊ .998    ÊÊÊ & ÊÊÊ .998    ÊÊÊ & ÊÊÊ .999
\\
$p=4$
& .771    ÊÊÊ & ÊÊÊ .884    ÊÊÊ & ÊÊÊ .944    ÊÊÊ & ÊÊÊ .973    ÊÊÊ & ÊÊÊ .988    ÊÊÊ & ÊÊÊ .994
\\
$p=8$  &  .273   ÊÊÊ & ÊÊÊ .546    ÊÊÊ & ÊÊÊ .749    ÊÊÊ & ÊÊÊ .891    ÊÊÊ & ÊÊÊ .946    ÊÊÊ & ÊÊÊ .973
\\
$p=25$ &  * &   .001   & ÊÊÊ .037    ÊÊÊ & ÊÊÊ .288    ÊÊÊ & ÊÊÊ .546    ÊÊÊ & ÊÊÊ .736
\\
$p=50$   & * & * & ÊÊÊ .000    ÊÊÊ & ÊÊÊ .005    ÊÊÊ & ÊÊÊ .079    ÊÊÊ & ÊÊÊ .289
\\
\hline
\end{tabular}
\end{center}

\vspace{5mm}

The comparison of Tables \ref{TablaValoresk2} and \ref{TableCovar}
shows that the 95\% percentiles shown in Table \ref{TablaValoresk2}
for the standard Gaussian distribution increase while the mean of
the determinant of the covariance matrix in Table \ref{TableCovar}
is above (roughly speaking) .750 and starts to decrease when the
determinant is below this quantity. The same behavior can be
observed in the Double Exponential case.

With respect to the Cauchy case, the mean of the determinant of the
employed estimate of the dispersion matrix never  falls below this
threshold, for the considered sample sizes and dimensions.

Precisely, this difference in the behavior between the Cauchy and
the other distributions makes some differences appear if we compare
the results between distributions with the same sample size and
dimension, but, in most cases, those differences are not
significantly large. Except when dimension is $p=50,$ the rate
between the largest and the lower 95\% percentiles lies between 1
and 1.5. However, it is noticeable that, if for each sample size and
distribution, we take the largest 95\% percentile along dimensions,
we obtain that the differences between the largest and the lower values are
just 1 except for sample size $n=1,000$ in which this difference is
2.

Thus, we propose to choose $k$, for a fixed dimension, as the
maximum of the 95\% percentiles along distributions and sample
sizes. However, if we prefer to fix the sample size, then we propose
to take the maximum on dimensions and distributions. Either way, to
have a full guarantee it is only required to take the maximum in the
table which gives the (surprisingly low) value $k=36$.

In this point the initial Hand's phrase is in force. There is no
doubt that, theoretically, the  accuracy of  the depth improves if
$k$ increases. However, for a {\it fixed sample size}, the  noise
coming from the sampling process makes large values for $k$ useless.
This point is reinforced in the following subsection where we
compute $D_T$, with $p=2$, using 1,000 vectors with no practical
gain.

\subsection{Testing homogeneity}
Our goal in this subsection is to show that the values obtained for
$k$ in Table  \ref{TablaValoresk2} give depths which provide results
similar to those obtained in practice with the Tukey depth. To this
end, we are going to reproduce the simulation study carried out in
\cite{Liu06}, where the authors apply  depth measures to test
differences in homogeneity between two distributions. Let us begin
by giving a brief description of the problem and the procedure.
Additional details can be found in \cite{Liu06}.

Assume that we have two random samples $\{X_1,...,X_{n_1}\}$ and
$\{Y_1,...,Y_{n_2}\}$ taken from the centered distributions $P $ and
$Q$ respectively. Let us assume that those distributions coincide
except for a scale factor, i.e., we are assuming that there exists
$r>0$ such that the r.v.'s   $\{rX_1,...,rX_{n_1}\}$ and
$\{Y_1,...,Y_{n_2}\}$ are identically distributed. The problem
consists in testing the hypotheses:
\begin{eqnarray*}
&H_0: &  r=1 \mbox{ (both scales are the same)}
\\
&H_a:& r >1 \ \mbox{Ê($Q$ has a larger scale)}.
\end{eqnarray*}

The idea is that, under the alternative, the observations in the
second sample should appear in the outside part of the joint sample
$ \{X_1,...,X_{n_1},Y_1,...,Y_{n_2}\}$, and, consequently, should
have lower depths than the points in the first sample. Thus, it is
possible to test $H_0$ against $H_a$ by computing the depths of the
points $ \{Y_1,...,Y_{n_2}\}$ in the joint sample, replacing them by
their ranks and rejecting $H_0$ if those ranks are small.

The Wilcoxon rank-sum test can be used to test when the ranks of the
points $ \{Y_1,...,Y_{n_2}\}$ are small. In \cite{Liu06} several
possibilities to break the ties are proposed. We have tried all of
them, with no important differences. Thus, we have selected to break
the ties at random as the only method to be shown here.

In Table \ref{TablaLiu1} we show the rate of rejections under the
exposed conditions  when we carry out the test at the significance
level $\alpha = .05$. The table also includes, between parenthesis,
the rejection rates when the random depth is replaced by the Tukey
depth using 1,000 directions uniformly scattered on the upper
halfspace.

The distributions used in the simulations are the 2-dimensional
standard Gaussian, and the double exponential and Cauchy with
independent marginals. We have centered the samples from the
Gaussian distribution in mean and the samples from the double
exponential and the Cauchy distributions in component-wise median.
We have considered the values $r=1, 1.2, 2$, and $n_1=n_2=n$ with $n
\in \{20,30,100\}$. We have done 10,000 simulations for each
combination of distribution, sample size and $r$.

Concerning the value of $k$ for the random depth, since we have to
compute random depths in  samples with sizes $2n \in \{40, 60,
200\}$ we have chosen $k=6,7$ and $8$ respectively. Those values are
close to the suggestions in Table \ref{TablaValoresk2} for $p=2$ and
the corresponding sample sizes.  We have not followed the hints at
the end of the previous section because we are interested in seeing
the behavior of the procedure with $k$ as low as possible.

\begin{Tabla} \label{TablaLiu1}
Rate of rejections in 10,000 simulations using $D_{T,k}$  with $k$
as shown (between  parenthesis, the rate with $D_T$) for the
considered distributions, sample sizes and values of $r$. Dimension
is $p=2$. The significance level is $.05$.
\end{Tabla}

\begin{tabular}{cc|ccc}
&& \multicolumn{3}{c}{Distribution}
\\
[-2mm]
Sample size & \multicolumn{1}{c}{Scale factor} &&&
\\
[-3mm]
&& \phantom{aaaa}Cauchy\phantom{aaaa}& \phantom{aa}Gaussian\phantom{aa} & D. exponential
\\
\hline $n=20$ &  $r=1$ & .054 \  \ (.057) & .059 \  \  (.061) & .053
\  \ (.056)
\\
$k=6$ & $r=1.2$ & .125 \  \ (.125) & .249 \  \  (.259)& .174 \  \
(.177)
\\
& $r=2$ &  .556 \  \  (.552) & .963 \  \  (.963) & .833 \  \ (.824)
\\
[1mm] \hline $n=30$ &  $r=1$ & .051 \  \  (.048) & .051 \  \ (.052)
& .052 \  \ (.053)
\\
$k=7$& $r=1.2$ & .140 \  \ (.148) & .316 \  \  (.325)& .219 \  \
(.214)
\\
& $r=2$ &  .691 \  \  (.699) & .995 \  \  (.996) & .940 \  \ (.941)
\\
[1mm] \hline $n=100$ &  $r=1$ & .086 \  \ (.055) & .055 \  \ (.057)
& .050 \  \ (.051)
\\
$k=8$& $r=1.2$ & .297 \  \ (.300) & .719 \  \  (.720)& .507 \  \
(.514)
\\
& $ r= 2$ &  .991 \  \  (.994) & 1 \  \  (1) & 1 \  \  (1)
\\
[1mm]
\hline
\end{tabular}

\vspace{5mm}

In \cite{Liu06} previous ideas are also applied to check the homogeneity between $K$ samples, $K>2$. The problem is the following. Let $\{ X_{1,1},..., X_{1,n_1}\},...,\{ X_{K,1},..., X_{K,n_K}\}$ be random samples obtained, respectively, from the distributions $P_1,...,P_K$ and let us assume that there exist $r_1,...,r_{K-1} >0$ such that the random vectors
$ r_1 X_{1,1},..., r_1X_{1,n_1}$,..., $r_{K-1} X_{K-1,1},..., r_{K-1}X_{K-1,n_{K-1}}$, $ X_{K,1},... , X_{K,n_K}$ are identically distributed.

We are interested in testing the following hypotheses:
\begin{eqnarray*}
&H_0: &  r_i=1 , i=1,...,K-1 \mbox{ (all scales are the same)}
\\
&H_a:& \mbox{ there exists } r _i \neq 1 \ \mbox{Ê(scales are
different)}.
\end{eqnarray*}

If we center separately each sample, join all the observations in a unique sample, compute the depths of all the points and transform those depths in ranks, then, we can apply the Kruskal-Wallis test to check if there are lacks of homogeneity between the ranks in each sub-sample.

We have carried out a simulation study applying previous procedure
to the Tukey depth and to the random Tukey depth in the
2-dimensional case with Gaussian distributions, $K=3$ and sample
sizes $n_1=n_2=n_3=n$, where $n \in \{20,30\}$. We have carried out
10,000 replications in each case at the significance level $\alpha =
.05$.

Concerning the selection of $k$, we have to compute the depths of points in samples with sizes $3n = 60,90$. Thus, according to Table \ref{TablaValoresk2}, we  have taken $k=7$ and $8$ random directions to project.

Results are shown in Table \ref{TablaLiu2}, where we also include
between parenthesis the results applying the same procedure with the
Tukey depth.

\begin{Tabla} \label{TablaLiu2}
Rate of rejections in 10,000 simulations using $D_{T,k}$  with $k$ as
shown (between parenthesis  the rate  with $D_T$) to test the
homogeneity in three samples  of Gaussian   distributions with
independent, identically distributed marginals and the exposed
values of $r$. Dimension is $p=2$.  The significance level is $.05$.

\end{Tabla}
\begin{center}
\begin{tabular}{c|cc}
& \multicolumn{2}{c}{Sample sizes and random directions}
\\
[-3mm]
\multicolumn{1}{c}{Covariance matrices} &&
\\
[-2mm]
& $n=20$ and $k=7$ \ & \ $n=30$ and $k=8$
\\
\hline $r_1=r_2 =1$ & .05 \ \ (.05) & .05 \ \ (.06)
\\
$r_1=r_2 =1.2$ & .16 \ \ (.15) &  .21 \ \ (.21)
\\
$r_1=2$, $r_2 =1.2$ & .89 \ \ (.89) &  .98 \ \ (.98)
\\
$r_1 =  r_2=2 $
& .96 \ \ (.97) &  1 \ \ (1)
\\
\hline
\end{tabular}
\end{center}

\vspace{5mm}

The results of both studies in this subsection are quite
encouraging, because there are no important differences among the
rejection rates with both  depths in spite of the big differences on
the employed number of directions.

\subsection{Computational time}
We end this section paying some attention to the required
computational time to compute the random Tukey depth. As a
comparison we have selected the time to compute the Mahalanobis
depth which is one of the quickest depths according to Table 1 in
\cite{Mosler}.

In Table \ref{TablaTiempos} we present the mean time, along 200
simulations,  employed to compute the random Tukey and Mahalanobis
depths for all points in a sample with the shown sizes and
dimensions. The number of employed random directions correspond with
those obtained in Table  \ref {TablaValoresk2}.

Since the random Tukey and Mahalanobis depths are computed on the
same samples, the first depth to be computed may have an advantage
in that the RAM memory may be cleaner than when the second one is
computed. In order to avoid this, we have computed the random Tukey
depth first 100 times and the Mahalanobis depth first 100 times.

The computations have been carried out on a computer Xserve G5, PowerPC G5 Dual 2.3 GHz and 2Gb of RAM memory.

\begin{Tabla} \label{TablaTiempos}
Time, in seconds, to compute the random Tukey and the Mahalanobis
depths (between parenthesis) of all points in a sample with size $n$
taken from a standard Gaussian distribution.
\end{Tabla}

\begin{center}
\begin{tabular}{cc|ccc}
& Random& \multicolumn{3}{c}{Sample size}
\\
[-3mm]
\multicolumn{1}{c}{Dimension} &&&&
\\
[-2mm]
& vectors& $n=100$ & $n=500$ & $n=1,000$
\\
\hline
$p=2$ & $k=8,9,11$ & 5.$9344\cdot10^{-4}$ \ (.0027)  & .0025\ (.0091) & .0060 \ (.0176)
\\
\hline
$p=4$ & $k=12,18,20$ & 8.$0011\cdot10^{-4}$ \ (.0026) & .0064\  (.0094)  & .0144 \  (.0178)
\\
\hline
$p=8$ & $k=13,27,35$
 & 8.$5459\cdot10^{-4}$ \  (.0027)  & .0119 \ (.0098)  & .0334 \ (.0185)
\\
\hline $p=25$ & $k=12,25,36$ & 8.$3209\cdot10^{-4}$  \  (.0031)  &
.0104 \  (.0111)   & .0356 \ (.0220)
\\
\hline
$p=50$ & $k=12,26,34$ & 8.$9296\cdot10^{-4}$ \  (.0048)  & .0116 \ (.0161)  & .0325 \  (.0296)
\\
\hline
\end{tabular}
\end{center}

\vspace{5mm}

The main computational effort to compute the Mahalanobis depth is devoted to obtain the inverse of the covariance matrix. In consequence, the computational time for this depth converges to infinity with the dimension.

On the other hand, the main difficulty in computing the random Tukey
depth is obtaining the projections of the involved points. Thus, the main
increment in required time for the random Tukey depth comes from the
increment in $k$. Taking into account that, according to Table  \ref
{TablaValoresk2}, $k$, as a function of $p$, is bounded, the
required time to compute random depths of a sample should not
increase as quickly. This is made apparent in Table
\ref{TablaTiempos} where, except for $n=100$, the maximum
computation time is not attained in the highest dimension.

\section{Functional  random Tukey depth. Functional classification} \label{SecFunctional}
An interesting possibility of the random Tukey depth is that it can
be straightforward extended to functional spaces. The only
requirement of the main results in \cite{Cues07} is that the sample
space has to be a separable Hilbert space. Thus, in this section we
will assume that we are considering a distribution \Prob \ defined
on this kind of space.

Concerning the number of random directions to take, it is possible
to consider the infinite dimensional Hilbert spaces as the limit of
finite dimensional euclidean spaces, and, then, given a sample size
$n$, it is enough to take $k$ as the maximum of the values provided
in Table \ref {TablaValoresk2} for this sample size.

For the reasons given in the introduction, we will directly check
how this depth works in practice. To this end, we have repeated the
classification problem  carried out in \cite{Romo06}, where the
authors handle a data set consisting of the growth curves of a
sample of 39 boys and 54 girls, with the goal to classify them, by
sex, using just this information. Heights were measured at 31 times
in the period from one to eighteen years. The data were taken from
the file growth.zip, downloaded from
ftp://ego.psych.mcgill.ca/pub/ramsay/FDAfuns/Matlab. The data are
drawn in Figure \ref{Alturas}.

\begin{center}
 \includegraphics[width=8cm]{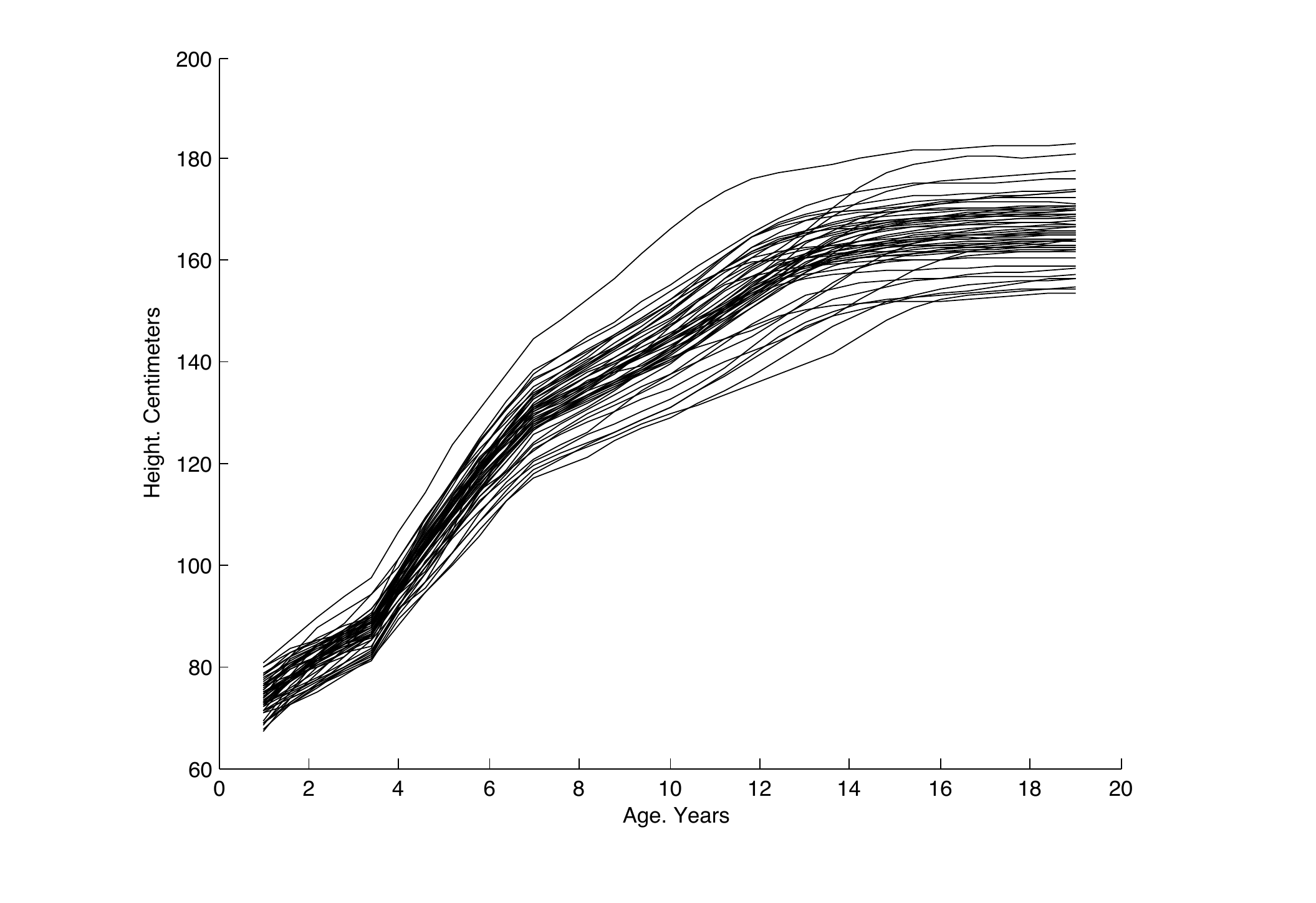}
  \hspace{-1.5cm}
 \includegraphics[width=8cm]{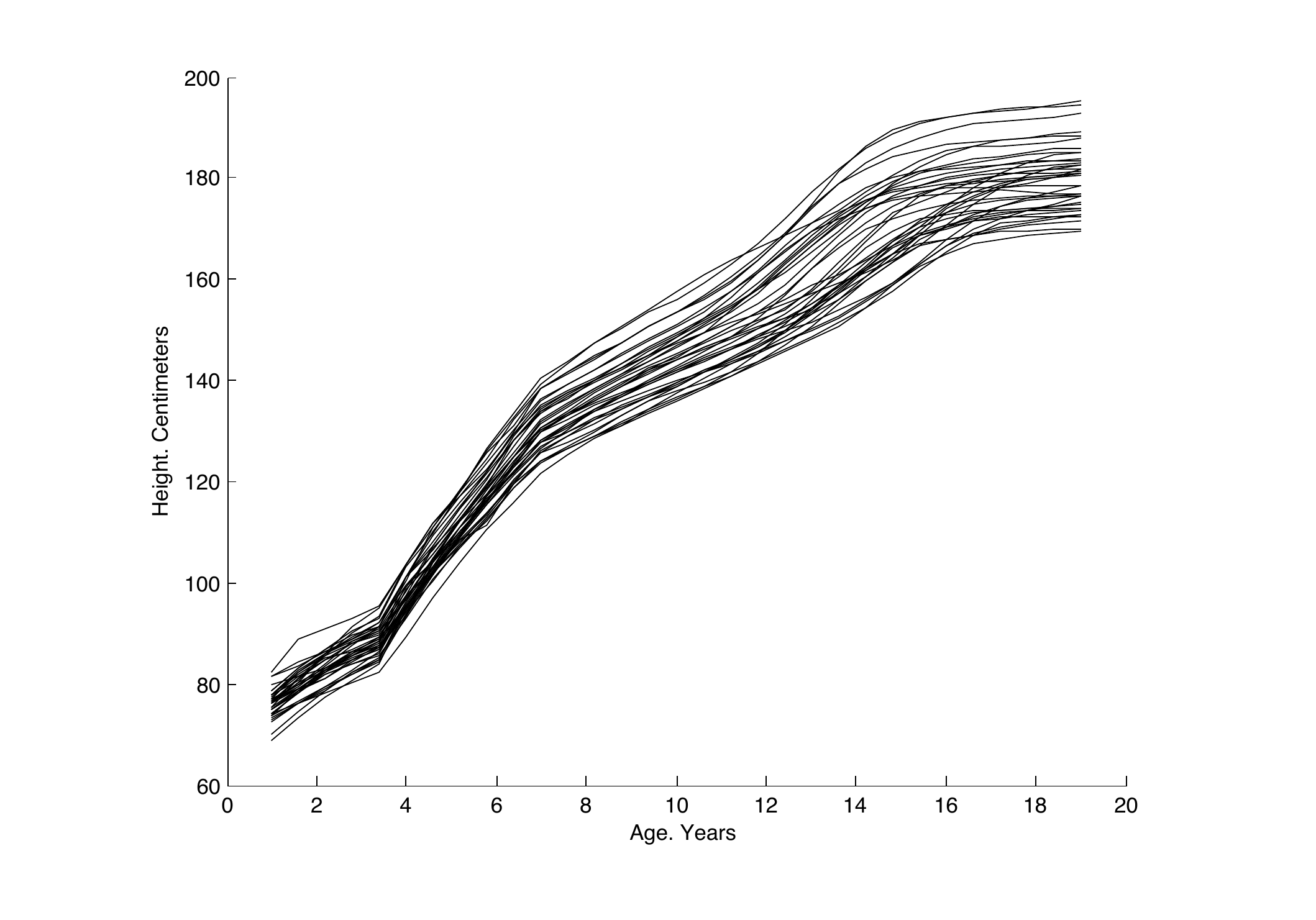}
\end{center}
\begin{Figu} \label{Alturas}
Growth curves of 54 girls (left hand side) and 39 boys (right hand side) measured 31 times each between 1 and 18 years.
\end{Figu}

It is well known that when handling this kind of data, it is useful
to consider not only the growth curve but also accelerations of
height (see, for instance, \cite{Rams}). However, since we are
mainly interested in comparing our results with those ones obtained
in \cite{Romo06}, where only growth curves were considered, here we
will do the same. Indeed,  we will repeat the study \cite{Romo06}
with three differences:

\begin{enumerate}

\item
Most importantly, we will replace the functional depths handled
there by the random Tukey depth.

\item
In \cite{Romo06} the authors consider the curves as elements in
$L^1[0,1]$, which is not possible here, because we need a separable
Hilbert space.

We will assume that \Hil \ is  the space of square-integrable functions in a given interval which, after re-scaling, we can assume to  be $[0,1]$. Thus, $\Hil =L^2[0,1]$ and given $f,g \in \Hil$ we have that $\langle f,g\rangle = \int_0^1 f(t)g(t)dt$.

\item
In \cite{Romo06}, the authors smoothed the original data using a spline basis. We have skipped this step because it did not seem necessary to us.

\end{enumerate}

The classification procedure can be extended to an arbitrary number
of groups, but, just to keep the notation as simple as possible, we
will assume that we have just two groups. Thus, let us assume that
we have two samples $X_1,...,X_n$ and $Y_1,....,Y_m$ in \Hil \
selected from two populations and that we are interested in
classifying another curve $Z \in \Hil$ in one of those groups using
a depth $D$ to be chosen later. Three classification methods are
proposed in \cite{Romo06}:

\vspace{4mm}
\noindent
{\bf 1.- Distance to the trimmed mean (M)}

Compute the depths of the points in the sample $X_1,...,X_n$.
Choose $\alpha \in [0,1)$.  The $\alpha$-trimmed mean of this sample, $\mu_{\alpha}(X)$, is  the mean of the $n\times (1-\alpha)$ deepest points.

Given $\beta \in [0,1)$, compute similarly $\mu_\beta(Y)$ the $\beta$-trimmed mean of the sample $Y_1,....,Y_m$.

Now, we classify $Z$ in the first group if
\[
\|Z - \mu_{\alpha}(X) \| < \|Z - \mu_\beta(Y) \| .
\]
Otherwise we classify $Z$ in the second group.

When applying  this method, $\alpha=\beta= .2.$

\vspace{4mm}
\noindent
{\bf 2.- Weighted average distance (AM)}

In some sense, in method M, each group is represented by its trimmed mean. Here,  we compute the distance between $Z$ and the group as a weighted mean of the distances between $Z$ and the members of the group where the weights are the depths of the points.

Thus,  we would classify the function $Z$ in the first group only if
\begin{equation}Ê\label{EqAM}
\frac{\sum_{i=1}^n  \|Z-X_i\| D_{X}(X_i)}{\sum_{i=1}^n  D_{X}(X_i)}
<
\frac{\sum_{j=1}^m  \|Z-Y_j\| D_{Y}(Y_j)}{\sum_{j=1}^n  D_{Y}(Y_j)},
\end{equation}
where  the subscripts  in $D_{X}$ and $D_{Y}$ mean that the depths are computed with respect to the empirical distribution associated to the corresponding sample.

\vspace{4mm}
\noindent
{\bf 3.- Trimmed weighted average distance (TAM)}

In the AM method, the result of the classification could be affected by the
number of elements in each sample if $n \neq m$. The solution for
this consists of taking a third value $l \leq \min (n,m)$ and
replacing (\ref{EqAM}) by
\[
\frac{\sum_{i=1}^l  \|Z-X_{(i)} \| D_{X}(X_{(i)} )}{\sum_{i=1}^l  D_{X}(X_{(i)} )}
<
\frac{\sum_{i=1}^l  \|Z-Y_{(i)} \| D_{Y}(Y_{(i)} )}{\sum_{i=1}^l  D_{Y}(Y_{(i)} )},
\]
where $X_{(1)}$ is the deepest point in the $X$-sample, $X_{(2)}$ is
the second  deepest point in the $X$-sample,... and similarly for
the $Y$-sample.

\vspace{5mm}

In \cite{Romo06} the authors consider three possibilities to split
the sample in training and validation sets. We have analyzed all
three possibilities, but in order to shorten the exposition we will
only present the results corresponding to the cross-validation
setting. However, we want to remark that, when using the random
Tukey depth, the differences between the error rates obtained with
those possibilities are less important than those reported in
\cite{Romo06}.

Regarding the selection of $k$, since the bigger sample size is
around $50$, following the suggestion at the beginning of this
section, we have taken $k=10$.

In Table \ref{TablaRomo} we show the obtained failure rates using
the described methods, the random Tukey depth and the depths
proposed in \cite{Romo06}. The last three columns contain the error
rates obtained with the depths handled  in \cite{Romo06}. They are
the band depth determined by three different curves (DS3), by four
different curves (DS4) and the generalized band depth (DGS). Their
values have been taken from Tables 1-3  in \cite{Romo06}.

On the other hand, taking into account the random nature of the
proposed depth, we have tried 10,000 times each classification method
with the random Tukey depth. The second column in Table
\ref{TablaRomo} contains the rate of errors we have obtained.

To facilitate comparisons, we present in bold the lowest
faliure rate for each method.

Once again, in spite of the low number of random projections, the
results are similar to those in \cite{Romo06}, the random Tukey
depth with the AM method being the global winner.

\begin{Tabla} \label{TablaRomo}
Rate of mistakes when classifying the growth curves by sex using
cross validation for the shown methods and depths.
\end{Tabla}

\begin{center}
\begin{tabular}{r|c|ccc}
Classification & Random Tukey & \multicolumn{3}{c}{Depths proposed in \cite{Romo06}}
\\
method &$k=10$ & DS3 & DS4 & DGS
\\
\hline
{M} & .2033   &\  .1828 \  & \ .1828  \     & \ {\bf .1613 } \
\\
\hline { AM} & {\bf  .1485}   & .2473   & .2473       & .1935
\\
\hline {TAM} & {\bf .1651}   & .2436   & .2436       & .1690
\\
\hline
\end{tabular}
\end{center}

\vspace{5mm}

\section{Discussion}
In this paper we introduce a random depth which can be considered as
a random approximation to the Tukey depth. The new depth is
interesting because of the little effort required into its
computation and because it can be extended to cover Hilbert valued
data.

From a theoretical point of view, this random depth enjoys no new
properties. Its interest lies in the fact that taking very few
one-dimensional projections, it is possible to obtain similar
results to those obtained with more involved depths. The number of
required projections is surprisingly low, indeed. In fact, for
samples sizes smaller or equal to $1,000,$ it seems that $36$
projections suffice for every dimension.

If the dimension of the space is fixed, the number of required
projections increases with the sample size. This dependence is
related to the fact that when the sample size is small, the
randomness included in the sample makes the gain achieved by
considering a high number of projections useless.

On the other hand, if we fix the sample size, the number of required
projections increases with the dimension until the point in which
the dimension is too large to allow a reasonable estimation of the
underlying distribution. From this point on, the number decreases.
The initial increment is related to the course of the
dimensionality. The later decrement is due to the uncertainty on the
parent distribution which makes a large number of projections
useless.

We consider remarkable the fact that a not too high number of
projections provides very good results even in the infinite
dimensional setting. This is shown in the comparisons  with some
other depths that we have carried out. Those studies, do not show
really important differences between the considered depths and the
random Tukey one. Thus, we conclude that, at least under considered
conditions, the random Tukey depth is an alternative which is worth
considering because of the small computational time required.

 \end{document}